\documentclass[apj]{emulateapj}
\usepackage{amsfonts,amsmath,graphicx,natbib,apjfonts,subfigure}

\def\sn{\mbox{SN\,2014J}}

\def\asec{\ifmmode ^{\prime\prime}\else$^{\prime\prime}$\fi}

\def\msun{\hbox{M$_{\odot}$}}

\def\msunyr{\mbox{\,${\rm M_{\odot}\, yr^{-1}}$}}
\def\mdot{\dot M}
\def\Mdot{\dot M}

\def\degs{\ifmmode ^{\circ}\else$^{\circ}$\fi}
\def\amin{\ifmmode ^{\prime}\else$^{\prime}$\fi}
\def\asec{\ifmmode ^{\prime\prime}\else$^{\prime\prime}$\fi}

\def\degs{\ifmmode ^{\circ}\else$^{\circ}$\fi}
\def\amin{\ifmmode ^{\prime}\else$^{\prime}$\fi}

\def\EE#1{\times 10^{#1}}

\def\cm{\mbox{\,cm}}

\def\cm3{\mbox{\,cm$^{-3}$}}
\def\kms{\mbox{\,km~s$^{-1}$}}

\def\kms{\mbox{\,km s$^{-1}$}}

\def\lsim{\!\!\!\phantom{\le}\smash{\buildrel{}\over
 {\lower2.5dd\hbox{$\buildrel{\lower2dd\hbox{$\displaystyle<$}}\over
                                 \sim$}}}\,\,}
\def\gsim{\!\!\!\phantom{\ge}\smash{\buildrel{}\over
{\lower2.5dd\hbox{$\buildrel{\lower2dd\hbox{$\displaystyle>$}}\over
                               \sim$}}}\,\,}
\def\iaa{1}
\def\cefca{2}
\def\unizar{3}
\def\albanova{4}
\def\oskar{5}
\def\jbca{6}
\def\jbo{7}
\def\jive{8}
\def\aao{9}
\def\uveg{10}
\def\donosti{11}
\def\oso{12}
\def\bonn{13}
\def\ouv{14}

\shorttitle{Constraints on the progenitor system and environs of SN~2014J}
\shortauthors{P\'erez-Torres et al.}

\begin{document}

\title{Constraints on the progenitor system and the environs of SN~2014J \\ 
from deep radio observations}

\author{M.~A. P\'erez-Torres  \altaffilmark{\iaa,\cefca,\unizar}, P. Lundqvist\altaffilmark{\albanova,\oskar}, 
R.~J. Beswick \altaffilmark{\jbca,\jbo}, C.~I. Bj\"ornsson\altaffilmark{\albanova}, T.W.B.~Muxlow \altaffilmark{\jbca,\jbo}, Z. Paragi \altaffilmark{\jive}, S.~Ryder \altaffilmark{\aao}, 
A. Alberdi \altaffilmark{\iaa}, C. Fransson \altaffilmark{\albanova,\oskar}, J.~M. Marcaide \altaffilmark{\uveg,\donosti}, 
I. Mart\'i-Vidal \altaffilmark{\oso}, E. Ros \altaffilmark{\bonn,\uveg,\ouv},
M.~K. Argo \altaffilmark{\jbca,\jbo}, J.~C. Guirado \altaffilmark{\uveg,\ouv}
}

\altaffiltext{\iaa}{Instituto de Astrof\'isica de Andaluc\'ia, Glorieta de las Astronom\'ia, s/n, E-18008 Granada, Spain.}
\altaffiltext{\cefca}{Centro de Estudios de la F\'isica del Cosmos de Arag\'on, E-44001 Teruel, Spain.}
\altaffiltext{\unizar}{Visiting Scientist: Departamento de F\'isica Teorica, Facultad de Ciencias, Universidad de Zaragoza, Spain.}
\altaffiltext{\albanova}{Department of Astronomy, AlbaNova University Center, Stockholm University, SE-10691 
             Stockholm, Sweden.}
\altaffiltext{\oskar}{The Oskar Klein Centre, AlbaNova, SE-10691 Stockholm, Sweden.}
\altaffiltext{\jbca}{Jodrell Bank Centre for Astrophysics, University of Manchester, Oxford Road, Manchester, M13 9PL, UK.}
\altaffiltext{\jbo}{Jodrell Bank Observatory, Macclesfield, Chesire, SK11 9DL, UK.}
\altaffiltext{\jive}{Joint Institute for VLBI in Europe, Postbus 2, 7990 AA Dwingeloo, NL.}
\altaffiltext{\aao}{Australian Astronomical Observatory, P.O. Box 915, North Ryde, NSW 1670, Australia.}
\altaffiltext{\uveg}{Departamento de Astronom\'ia i Astrof\'isica, Universidad de Valencia, E-46100 Burjassot, Valencia, Spain.}
\altaffiltext{\donosti}{Donosita International Physics Center, Paseo de Manuel de Lardizabal 4, 20018 Donostia-San Sebasti\'an, Spain.}
\altaffiltext{\oso}{Onsala Space Observatory, Chalmers University of Technology, SE-43992 Onsala, Sweden.}
\altaffiltext{\bonn}{Max-Planck-Institut f\"ur Radioastronomie, D-53121 Bonn, Germany.}
\altaffiltext{\ouv}{Observatorio Astron\'omico, Universidad de Valencia, E-46980 Paterna, Valencia, Spain.}

\begin{abstract}
 We report deep EVN and eMERLIN observations of the Type Ia SN~2014J in the nearby galaxy M~82. 
Our observations represent, together with JVLA observations of SNe~2011fe and 2014J, the most
sensitive radio studies of Type Ia SNe ever. By combining data
and a proper modeling of the radio emission, we constrain the mass-loss rate from the progenitor system of 
SN~2014J to $\dot{M} \lesssim 7.0\EE{-10}~\msunyr$ (for a wind speed of 
$100 \kms$). If the medium around the supernova is uniform, then $n_{\rm ISM} \lesssim 1.3 \cm3$, which is 
the most stringent limit for the (uniform) density around a Type Ia SN.
Our deep upper limits favor a double-degenerate (DD) scenario--involving two WD stars--for the progenitor system of SN~2014J, as such systems have less circumstellar gas than our upper limits.  By contrast, most single-degenerate (SD) scenarios, i.e., the wide family of progenitor systems where a red giant, main-sequence, or sub-giant star donates mass to a exploding WD, 
are ruled out by our observations\footnote{While completing our work, we noticed that a paper by
\citet{Margutti2014} was submitted to {\sl The Astrophysical Journal}. From a non-detection of X-ray
emission from SN 2014J, the authors obtain limits of $\dot{M} \lsim 1.2 \times 10^{-9}$ \msunyr\ (for a wind speed of 
$100 \kms$) and $n_{\rm ISM} \lsim 3.5 \cm3$, for the $\rho \propto r^{-2}$ wind and constant 
density cases, respectively. As these limits are less constraining than ours, the 
findings by \citet{Margutti2014} do not alter our conclusions. The X-ray results are, however, important 
to rule out free-free and synchrotron self-absorption as a reason for the radio non-detections.}.
Our estimates on the limits to the gas density surrounding SN2011fe, using the 
flux density limits from \citet{cho12}, agree well  with their results.
Although we discuss possibilities for a SD scenario to pass observational tests, as well as uncertainties
in the modeling of the radio emission,  the evidence from SNe~2011fe and 2014J points in the direction of a 
DD scenario for both.
\end{abstract}


\section{Introduction}
\label{sec:intro}

Type Ia supernovae (SNe) are the end-products of white dwarfs with a
mass approaching the Chandrasekhar limit, which results in a
thermonuclear explosion of the star.  In addition to their use as
cosmological distance indicators \citep[e.g.,][]{rie98,per99}, Type Ia SNe 
(henceforth SNe Ia) are a major contributor to the chemical
evolution of galaxies. It is therefore unfortunate that we do not yet
know what makes a SN Ia.  This lack of knowledge makes it difficult to
gain a physical understanding of the explosions, so that we can model
possible evolution, which compromises their use as distance
indicators. It also means we do not fully understand the timescale
over which SNe Ia turn on, adding a large uncertainty to our
understanding of the chemical evolution of galaxies.

Unveiling the progenitor scenario for SNe Ia is difficult because
white dwarfs (WDs) can, theoretically, reach their fatal Chandrasekhar mass
in many ways, and disentangling which is the correct one (if there
is just one), is challenging from an observational point of
view. Nonetheless, there are two basic families of models leading to a
SN Ia, the single-degenerate model (SD) and the double-degenerate
model (DD).  In the SD scenario, a WD accretes mass from a
hydrogen-rich  companion star before reaching a mass close to the
Chandrasekhar mass and going off as supernova.  In the DD scenario, two
WDs merge, with the more-massive WD being thought to tidally disrupt and accrete the lower-mass WD \citep[see, e.g.,][and references therein]{mao14}.

Observations can potentially discriminate between the progenitor
models of SNe Ia.  For example, in all scenarios with mass
transfer from a companion, a significant amount of circumstellar gas
is expected \citep[see, e.g.,][]{bra95}, and therefore a shock is
bound to form when the supernova ejecta are expelled. The situation
would then be very similar to circumstellar interaction in
core-collapse SNe, where the interaction of the blast wave from the
supernova with its circumstellar medium results in strong radio and
X-ray emission \citep{che82b}.  On the other hand, the
DD scenario will not give rise to any circumstellar medium close to the progenitor system,
and hence essentially no radio emission is expected.

Radio and X-ray observations of SN~2011fe have provided the most sensitive constraints on 
possible circumstellar material \citep{cho12,mar12} around a normal SN~Ia. The claimed limits on mass
loss rate from the progenitor system are $\mdot = 6\times10^{-10}$ \msunyr\ 
and $\mdot = 2\times10^{-9}$ \msunyr\ from 
radio \citep{cho12} and X-rays \citep{mar12}, respectively, assuming 
a wind velocity of 100~km~s$^{-1}$.  Radio \citep[e.g.,][]{panagia06,han11} 
and X-ray \citep[e.g.,][]{hug07,rus12} observations of other, more distant SNe~Ia, have resulted
in less constraining upper limits on wind density. The non-detections of radio and X-ray emission
from SNe~Ia have added to a growing consensus that  a large fraction of SNe~Ia may not be the 
result of SD scenarios \citep[e.g.,][]{mao14}. 

Despite the non-detection of radio and X-ray emission, there is evidence of possible circumstellar 
material in the form of time-varying absorption features in the optical Na~I~D line for a few SNe~Ia 
\citep{pat07,sim09,dil12}, supposed to arise in circumstellar shells. The exact location of the absorbing
gas is still debated \citep[e.g.,][]{chu08,soa14}, and probably varies from case to case. The number of 
SNe~Ia showing indications of circumstellar shells could be significant, although the uncertainty is still 
large ((18$\pm11$)\%; \citealt{ste14}). Just as with the radio and X-rays, no optical
circumstellar emission lines from normal SNe~Ia have yet been detected \citep[e.g.,][]{lun13}, although there
are a few cases with strong emission \citep[see, e.g.,][for an overview]{mao14}. Those
SNe~Ia with strong circumstellar interaction constitute a very small fraction of all SNe~Ia, probably only
$\sim 1$\% \citep{chu04}.

Recently, \citet{foss14} serendipitously discovered \sn\ in
the nearby galaxy M~82 (D=3.5 Mpc).  \citet{cao14} 
classified SN~2014J as a SN Ia, which makes it the closest SN Ia
since SN~1986G in Cen A, almost three decades ago. The supernova
exploded between UT 14.56 Jan 2014 and 15.57 Jan 2014 according to
the imaging obtained by \citet{ita14}\footnote{see http://www.k-itagaki.jp/psn-m82.jpg}, 
and its J2000.0 coordinates are RA=09:55:42.121, Dec=+69:40:25.88 \citep{smi14}. 
For a further discussion on the discovery and early rise of the optical/IR emission, we refer
to \citet{goo14} and \citet{zheng14}.
The vicinity of \sn\ makes it a unique case for probing its prompt radio
emission, and thus constrain its progenitor system.

\section{Observations and data reduction}
\label{sec:obs}

We observed SN2014J with the electronic Multi Element Radio
Interferometric Network (eMERLIN) at 1.55 and 6.17 GHz, and with the electronic European Very
Long Baseline Interferometry Network (EVN) at a frequency of 1.66 GHz.
We show in Table 1 the summary for our observations, along with radio data  obtained by others.

\subsection{eMERLIN observations}

We observed \sn\ with eMERLIN on 28 January 2014, at a frequency of
1.55 GHz, and on 29-30 January 2014, at a
frequency of 6.17 GHz.  Our observing array included, at both
frequencies, all eMERLIN stations (Lovell, Jodrell Mk2, Darham, Pickmere, Cambridge, Defford, Knockin).
Given the expected faintness of \sn\, we used a phase-reference observing scheme, with
$\sim$8 minutes spent on the SN, and $\sim$2 minutes on the nearby, bright
phase-calibrator J0955+6903 (RA=09:55:33.1731; Dec=69:03:55.061). We
used 3C286 as our absolute flux density calibrator, and OQ208 as
bandpass calibrator.  We observed in dual-polarization mode at both
frequencies.  The bandwidth at 1.55 (6.17) GHz was of 512 (1024)
MHz. Each of those frequency bands was split into 4 (8)
spectral windows (SPW) of 128 MHz each. 
Each SPW was in turn split into 512 channels/polarisation.

\begin{figure*}
\centering
\includegraphics[width=18cm,angle=0]{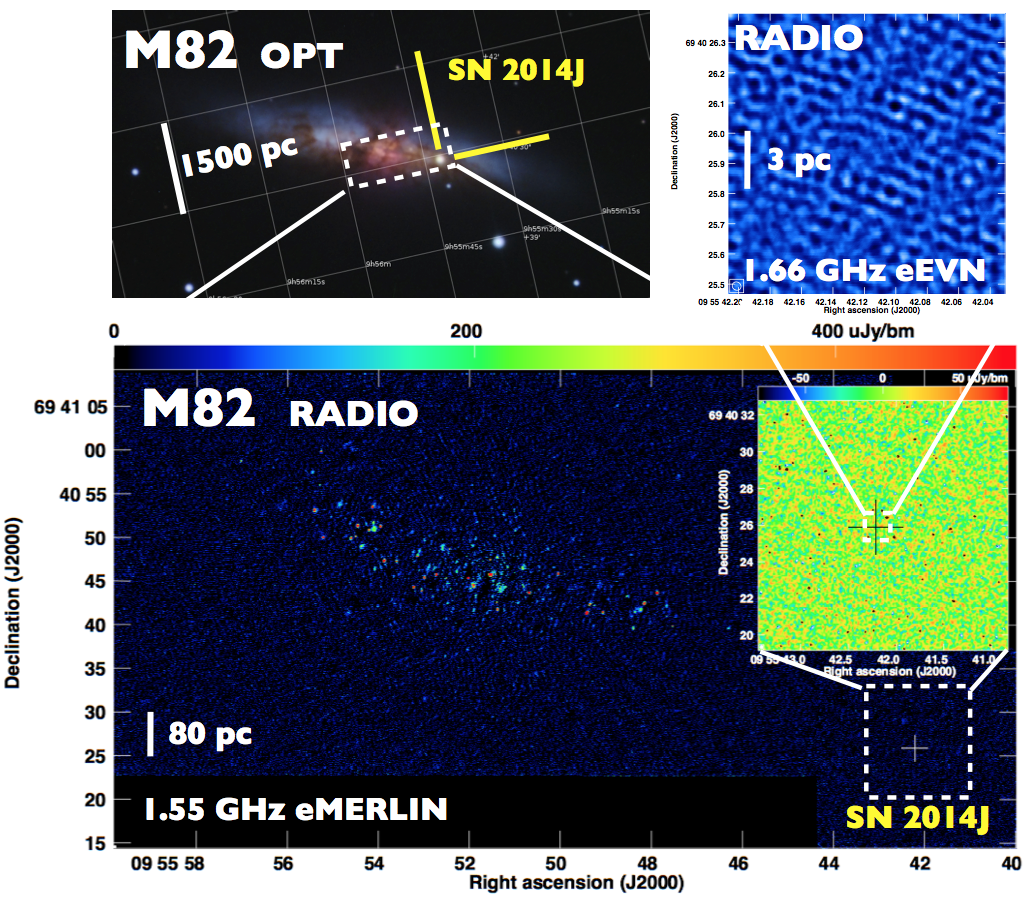}
\caption{\emph{Top left:} RGB optical image of  the very nearby galaxy M~82 and its \sn\, obtained with the
50~cm telescope of the Observatorio Astron\'omico de Aras, Spain, on 31 January 2014.  \emph{Bottom:} 1.55 GHz eMERLIN radio image of M~82 from our observations on 28 January 2014, which shows a large number of supernovae and supernova remnants. The inset is a blow-up image of the region around SN~2014J. \emph{Top right:} 1.66 GHz eEVN image of the SN~2014J field.
 }
\label{fig:fig1}
\end{figure*}

\begin{table}
\caption{Log of radio observations}
\scalebox{0.9}{
\begin{tabular}{lcclccrc}
\tableline\tableline
Starting & $T$ & $t_{\rm int}$ & Array & $\nu$ & $S_\nu$ & L$_{\nu,23}$  & $\dot M_{-9}$ \cr
UT        &   day &  hours          &           & GHz  & $\mu$Jy &
   &  
 \cr
\tableline
%
 %
Jan 23.2  &    8.2 &  $-$     & JVLA        &   5.50   &   12.0   &   1.77      &  0.70~(4.2)      \cr           
Jan 24.4  &    9.4 &  $-$   & JVLA        &   22.0 &   24.0     &   3.51    &  3.7~(22)       \cr           
Jan 28.8  &  13.8 &  13.6   & eMERLIN &  1.55 & 37.2      &   5.46  &   1.15~(7.0)     \cr
Jan 29.5  &  14.5 &  14.0   & eMERLIN &  6.17 & 40.8     &     5.97  & 3.6~(22)  \cr
Feb  4.0   &  20.0 &  11.0  & eEVN       &  1.66 & 32.4  & 4.74   & 1.69~(10)  \cr
Feb 19.1  &  35.0 &  10.0  & eEVN       &  1.66 &  28.5  & 4.17   & 2.9~(16)  \cr
\tableline
\end{tabular}
}
\tablecomments{The columns starting from left to right are as follows: starting observing time, UT; mean 
observing epoch (in days since explosion, assumed to be on Jan 15.0);  integration time, in hr;
 facility; central frequency in GHz; 3$\sigma$ flux density upper limits, in $\mu$Jy; 
 the corresponding 3$\sigma$ spectral luminosity,  assuming a distance of 3.5 Mpc to M82, in units of 
 $10^{23}$\,erg\,s$^{-1}$\,Hz$^{-1}$; inferred 3$\sigma$upper limit to the  mass-loss rate in units of 
 $10^{-9} M_\odot\,{\rm yr}^{-1}$, for an  assumed wind velocity of $100 \kms$. 
  (The values for $\dot M_{-9}$ are for $\epsilon_{\rm B} = 0.1$ and, in parenthesis, for 
  $\epsilon_{\rm B} = 0.01$, and have been calculated using our model described in \S 4.) The Jansky 
  Very Large Array (JVLA) data are taken from \citet{Chandler2014}, while the rest of the data are from this work.}
 \label{tab:RadioLog}
\end{table}

We loaded the data into the NRAO Astronomical Image Processing System
(\emph{AIPS}) of the National Radio Astronomy Observatory (NRAO, USA),
after having averaged them to 128 channels per SPW (i.e., channel width
of 500 kHz).  We used AIPS for calibration, data inspection, and
flagging, using standard procedures.  We lost $\sim$15\% of the data
due to Radio Frequency Interference (RFI). We imaged 
the $\sim$13\arcmin ($\sim$4\arcmin) field of view of our 1.55 (6.17) GHz observations,
including M82, which has a strong and spatially complex radio
structure, using a robust=0 uv-weighting scheme.  We used those in-beam radio sources  to refine the target
 field calibration via several rounds of phase and amplitude
 self-calibration. Following these rounds of self-calibration, we
reweighted the target visibilities to account for difference in
the sensitivity of the individual e-MERLIN antennas.  Our final imaging
yielded 13.6 and 12.4 $\mu$Jy/bm r.m.s. noise levels at the location of
 SN2014J at 1.55 and 1.67 GHz, respectively.

\subsection{eEVN observations}

We observed our target source, \sn\ on 3-4 February 2014 and 19 February 2014, using the eEVN at 1.66\,GHz.  We used a
sustained data recording rate of 1024 Mbit s$^{-1}$, in
dual-polarisation mode and with 2-bit sampling. Each frequency band
was split into 8 intermediate subbands of 16 MHz bandwidth
each, for a total synthesized bandwidth of 128~MHz.  Each subband was in
turn split into 128 (64) spectral channels of 125 (250)~kHz bandwidth each for
the observations on 3-4 February (19 February) 2014.
 
Our observations on 3 February included the following six
antennas of the EVN: Effelsberg, Westerbork (phased array), Jodrell
Bank (Mk 2), Medicina, Onsala, and Torun.  In addition to these
antennas, our observing run on 19 February included also the antennas
of Noto and Sheshan.  We observed our target source, SN 2014J,
phase-referenced to the core of the nearby galaxy M81, known to be
very compact at VLBI scales, with a typical duty cycle of 5 minutes.
We used the strong source DA193 as fringe finder and bandpass
calibrator.  All the data were correlated at the EVN MkIV data
processor of the Joint Institute for VLBI in Europe (JIVE, the
Netherlands), using an averaging time of 1~s.
 
We used \emph{AIPS}  for calibration,
data inspection, and flagging of our eEVN data, using standard procedures.  Those steps
included a-priori gain calibration (using the measured gains and
system temperatures of each antenna), parallactic angle correction and
correction for ionosphere effects.  We then aligned the visibility
phases in the different subbands, i.e., ``fringe-fitted'' the data, solved for the residual delays and delay rates, 
and interpolated the resulting gains into the scans of \sn.  We
then imaged a field of view of 3\arcsec$\times$3\arcsec centered at the position
given by \citet{smi14}, and applied standard imaging procedures using
\emph{AIPS}, without averaging the data either in time, or frequency,
to prevent time- and bandwidth smearing of the images. We used natural uv-weighting to maximize the signal to noise ratio in our final images.

\section{A model for the radio emission from Type Ia SNe}
\label{sec:model}

The radio and X-ray non-detections of SNe~Ia, in conjunction with indications of circumstellar shells around 
some SNe~Ia (see \S\ref{sec:intro}), is a conundrum that yet has to find a solution. The nearby northern hemisphere 
SNe~2011fe and  2014J offer a possibility to use the most sensitive radio facilities present to probe 
circumstellar emission. In particular, we now interpret the upper limits on radio emission from SN~2014J 
in \S\ref{sec:obs}  within the framework of circumstellar interaction. Indeed, when the supernova shock-wave ploughs
through the circumstellar gas, a high-energy density shell forms. Within this shell, electrons are accelerated to
relativistic speeds and significant magnetic fields are generated, especially if the circumstellar gas is 
pre-ionized. For the low wind densities discussed in this paper, pre-ionization is likely to occur \citep{cum96}. 
The relativistic electrons radiate synchrotron (radio) emission \citep[e.g.,][]{che82b}.

A proper modeling of the radio emission from SNe requires, in principle, 
taking into account Coulomb, synchrotron, and (inverse) Compton losses of the relativistic 
electrons. However, since we only have upper limits for the radio emission from SN~2014J, we 
will discuss the radio emission from SNe~Ia within a scenario of Type Ib/c 
SNe \citep[see, e.g.,][]{che06}, neglecting energy losses for the 
relativistic electrons \citep[c.f.][for a more general treatment]{fra98,vidal11}. 
The spectrum of the radio emission from those SNe follows the ``Synchrotron
Self-Absorption'' (SSA) form, i.e., a rising power law with $\nu^{5/2}$ (low-frequency, optically 
thick regime), and a declining power law, $\nu^\alpha$ (high frequency, optically thin regime), 
where $\alpha$ is assumed to be constant. For most well studied SNe, $\alpha\approx -1$ \citep{che06}.
We assume that electrons are 
accelerated to relativistic energies, with a power law distribution, $dN/dE = N_0E^{-p}$; 
where $E=\gamma m_ec^2$ is the energy of the electrons and $\gamma$ is the Lorentz factor.
For synchrotron emission, $\alpha = (p-1)/2$, which indicates that $p\approx3$ should  be used.
 
Here, we study both the case of a circumstellar structure created by a wind, as well as the case with constant density circumstellar gas. For the wind case, we make the standard assumption that the SN progenitor has been losing matter at a constant rate, $\dot M$, so that the circumstellar density
has a radial profile: $\rho(r) = n_{\rm CSM}(r)\mu = \dot M/(4\pi r^2 v_w)$, where
$v_w$ is the wind velocity, $r$ is the radial distance from the star,
$n(r)$ is the particle density and $\mu$ is the mean atomic weight of the
circumstellar matter.

To calculate the shock expansion, we use the thin-shell approximation
\citep{che82b}, with the extensions of \citet{tru99}.
We assume that the innermost ejecta has a density slope 
of $\rho_{\rm ej,inner} \propto r^{-\delta}$, which at some velocity of the ejecta rolls over 
to a  steeper density profile, $\rho_{\rm ej,outer} \propto r^{-n}$ ($n > \delta$). 
We assume $\delta = -2$, motivated by the explosion models of \citet{fink14}, and 
use $n = 10.2$, which is a good approximation  to the 
outer density profile of a supernova that stems from a radiative star \citep{matzner99}.
Assuming an ejecta mass of $1.4~\msun$ and a kinetic energy of the explosion
of 10$^{51}$ erg, the break in power-law index in our model occurs at $\approx 1.25\EE4 \kms$, which
agrees with the angle-averaged results of \citet{fink14}.
The supernova expansion can be well approximated by a power law, 
$r_s \propto t^m$, so that the shock speed, $v_s = m \, r_s/t$ \citep{che82b}. Here, $m = (n-3)/(n-s)$,
and $s$ is the density slope of the circumstellar gas, which for the steady wind case is $s=2$. 
(In \S\ref{sec:ism} we also discuss the case $s=0$.) 
The shock speed at 10 days in this model 
is  $v_s \approx 8.3 \EE4 \kms$ for $\mdot = 1 \times10^{-9}$ \msunyr\  
and $v_w = 100 \kms$; $v_s$ and $r_s$ both scale as $(\dot M/v_w)^{-1/(n-s)}$.

For any sensible pre-supernova wind speed, the
supernova shock is strong. Assuming a polytropic gas with $\gamma =
5/3$,  the compression of the gas across the shock is $\eta = 4$, and the post-shock thermal energy
 density is $u_{\rm th} = \frac{9}{8}\rho v_s^2$, where $\rho$ is the pre-shock density.
Following \citet{che06}, we denote $\epsilon_{\rm B} = u_{\rm
  B} / u_{\rm th}$, where $u_{\rm B} = B^2/(8\pi)$ is the (post-shock)
magnetic energy density; and $\epsilon_{\rm rel} = u_{\rm  rel} /
u_{\rm th} $, where $u_{\rm  rel}$ is the energy density of the relativistic particles, assumed for 
simplicity to be electrons.

We assume that the power law index of the relativistic
electron population stays constant with time at $p = 3$, although we 
have also studied cases with $p = 2.5$ (see \S\ref{sec:sens-res}). 

The most uncertain parameters refer to the microphysics of the
shocked gas, namely $\epsilon_{\rm rel}$ and, to a greater extent,
$\epsilon_{\rm B}$. Indeed, it seems that $\epsilon_{\rm rel} \sim 0.1$ with some small dispersion 
around this value \citep{che06}, whereas 
$\epsilon_{\rm B}$ appears to vary more among supernovae, and is hence
largely unknown. Therefore, we fix $\epsilon_{\rm rel} = 0.1$, 
and take $\epsilon_{\rm B}$ as a free parameter. 
We can easily find $N_0$ by integrating the relativistic electron
distribution between  $E_{\rm min} = \gamma_{\rm min}\,m_ec^2$ and infinity, which yields
$N_0 = (p-2) \epsilon_{\rm rel} u_{\rm th} E_{\rm min}^{p-2}$.

We estimate the minimum Lorentz factor of the relativistic electrons, $\gamma_{\rm min}$, 
assuming that all postshock electrons go into the power-law distribution with energy index $p$
 \citep[cf.][]{che06}. This means that  $\epsilon_{\rm rel} u_{\rm th} \approx \eta n_e E_{\rm min} [(p-1)/(p-2)]$
\citep[see also][]{cho12}. Here, $n_{\rm e}$ is the electron density of the pre-shocked gas when it is 
fully ionized. We  assume a mix of H and 
He with an abundance ratio 10:1, which together with $\epsilon_{\rm rel} = 0.1$ and $\eta = 4$ means that 
$\gamma_{\rm min} \approx 1.64~[v_s /(70\,000\kms)]^2$. Following \citet{che98}, we add
the  constraint that $\gamma_{\rm min}  \geq 1$.

To calculate the synchrotron spectrum, we follow the method by \citet{bjo14}, i.e.,
we use the observational evidence that the brightness temperature, $T_{\rm bright}$, is expected to 
be somewhat below 10$^{11}$~K \citep[cf.][]{red94,bjo14}. While we defer a more complete
discussion about this to a future paper (C.-I. Bj\"ornsson, in preparation), we have chosen a likely value of 
$T_{\rm bright} = 5\EE{10}$~K, which should be correct to within a factor of $\sim 2$. The intensity at the 
frequency of the peak of the synchrotron spectrum, ${\nu}_{\rm peak}$,  is then defined 
as $I_{{\nu}_{\rm peak}} \equiv 2kT_{\rm bright} ({\nu}_{\rm peak}/c)^2$,
whereas the intensity at any frequency is $I_{\nu} = S_{\nu} [1 - {\rm exp}(-{\tau_{\nu}})]$. Here
$S_{\nu} \propto \nu^{5/2}$ is the source function and $\tau_{\nu}$ the synchrotron optical depth.
The latter is just $\tau_{\nu} =  \kappa_{\nu}\Delta s$, where $\Delta s$ is the path length through 
the emitting region along the line of sight, and $\kappa_{\nu} = \varkappa(p) N_0 B^{(p+2)/2} \nu^{-(p+4)/2}$.
Like  \citet{che98}, we make the simplification that $B\ {\rm sin}(\theta) \approx B$, where $\theta$ is the 
particle pitch angle. The constant $\varkappa(p)$ can be found in, e.g., \citet{ryb79}. The path
length $\Delta s$ depends on the thickness of the synchrotron emitting region, $\Delta r$. At the center,
$\xi_h  \equiv \Delta s (h)/ (2 \Delta r) = 1$ (assuming the supernova ejecta to be transparent to radio 
emission), but can become significantly larger than unity toward the limb. $h$ is the normalized impact 
parameter, so that $0 \leq h \leq 1$. We assume constant properties of the plasma 
within $\Delta r$. For $s=2$, we have assumed a thickness of $\Delta r/r_s = 0.2$, which corresponds to 
that of the shocked circumstellar gas for $n \sim 10$ and $s = 2$ in the similarity solutions 
of \citet{che82a}, namely, $\Delta r/r_s \simeq 0.19\pm0.01$ for $9 \leq n \leq 12$. For $s = 0$, the 
similarity solutions give $\Delta r/r_s \simeq 0.116\pm0.008$ for the same range in $n$, and 
we have chosen $\Delta r / r_s \approx 0.12$ for this $s$-value. $\xi_h$ is therefore just due to the 
geometrical increase of the path length as $h$ increases.

For convenience, we introduce, in addition to ${\nu}_{\rm peak}$, also the frequency ${\nu}_{\rm abs}$, 
defined as $\tau_{\nu_{\rm abs}} = 1$. In general, $\tau_{\nu} = (\nu / {\nu}_{\rm abs})^{-(p+4)/2}$. 
For $h = 0$, we denote
$\nu_{\rm abs} = \nu_{{\rm abs},0}$, $\tau_{\nu_{\rm abs}} = \tau_{\nu_{{\rm abs},0}}$ and $\tau_{\nu} = \tau_{\nu_{0}}$. We can
then derive the intensity for any impact parameter as

\begin{equation}
	I_{\nu}\left(h\right)=\frac{2kT_{\rm bright}} {c^2 f\left(\frac{{\nu}_{\rm peak}}{{\nu}_{\rm abs}}\right)}
	                 \frac{\nu^{5/2}} {\nu_{{\rm abs},0}^{1/2}}
	                 \left[
	                 1-{\rm exp}\left(-\xi_h \tau_{\nu_{0}}\right)
	                 \right],
\label{eq:Intensity}
\end{equation}
where $f(x)$ depends on $p$ such that

\begin{equation}
	f (x) = x^{1/2} \left[1 - {\rm exp} \left( - x^{-\left(p+4\right)/2} \right) \right]
\label{eq:fx}
\end{equation}

\citep[see also][]{bjo14}. 
For $p=3~(2.5)$, $\nu_{\rm peak}/\nu_{\rm abs} \approx 1.137~(1.235)$ and $f(x) \approx 0.503~(0.440)$. 
To obtain the luminosity, one integrates over $h$, so 
that  $L_{\nu} = 8 \pi^2 r_{s}^{2} \int_0^1 I_{\nu}(h) h dh$. The longer path length toward the limb makes
$L_{\nu}$ larger for the optically thin part of the spectrum than just assuming 
$L_{\nu} = L_{{\nu},0}  = 4 \pi^2 r_{s}^{2} I_{\nu}(0)$. For $p=3$, 
the factor $\vartheta_{\nu}  \equiv L_{\nu} / L_{{\nu},0}$ in the
optically thin part is a weak function of $\Delta r$, being $\approx 1.81~(1.63)$ for  $\Delta r / r_s = 0.1~(0.2)$.
For the  optically thick part, $\vartheta_{\nu}  = 1$. This makes the {\it observed} spectrum peak at 
a somewhat higher frequency than ${\nu}_{\rm peak}$.

\begin{figure}
\centering
\includegraphics[width=9cm,angle=0]{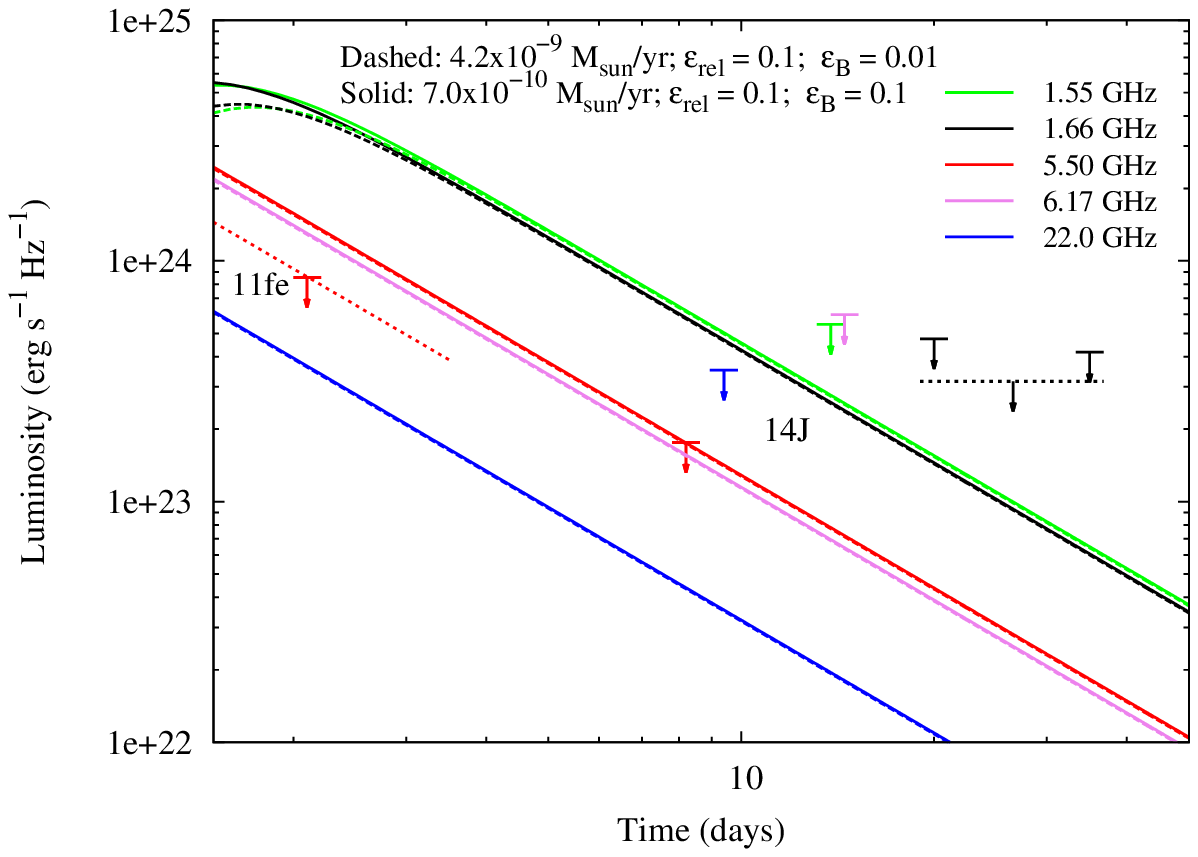}
\caption{Predicted radio light curves of \sn\ in M 82 for an assumed mass-loss rate of $\dot M = 7.0\times10^{-10}$\msunyr (solid lines), and for  $\dot M = 4.2\times10^{-9}$\msunyr (dashed lines). For the former we used $\epsilon_{\rm B} = 0.1$ and for the latter 
$\epsilon_{\rm B} = 0.01$.  The data points (cf. Table ~\ref{tab:RadioLog}) with 3$\sigma$ upper limits for \sn\ are in the right part of the figure. Shown in the figure is also the earliest 5.9 GHz 3$\sigma$ upper  limit for SN~2011fe \citep{cho12}, scaled to its distance of 6.4 Mpc, together with a dotted line marking the predicted evolution for $\dot M = 5.0\times10^{-10}$\msunyr (for $\epsilon_{\rm B} = 0.1$). Common parameters in all models are $\epsilon_{\rm rel} = 0.1$, $p=3$ and $v_w = 100$\kms.  See text for further details.
}
\label{fig:Lcurves}
\end{figure}

\section{Results}
\label{sec:results}

\subsection{Modeling the data for SN~2014J}
The radio emission from the supernova is subject to both SSA and possible 
external free-free absorption in the ambient medium. While we include SSA in our model, we do not include free-free absorption since, as we show below, it is negligible. In previous analyses of SNe~Ia \citep{panagia06,han11}, 
free-free absorption was assumed to
be the most important factor to derive wind densities. For SNe~2011fe and 2014J, X-ray non-detections
\citep{mar12,Margutti2014} have put limits on $\mdot/v_w$ of order $10^{-9} \msunyr$ for $v_w = 100 \kms$.
From Equation 6 in \citet{lun88} it follows that the free-free optical depth, $\tau_{\rm ff}$, for a fully ionized 
wind at $10^4$~K and moving  at $v_w = 100 \kms$, is $\tau_{\rm ff} \sim 10^{-8} \lambda^{2} (\Mdot/10^{-9} 
\msunyr)^2 (r_s/10^{15} {\rm cm})^{-3}$, where $\lambda$ is in cm. For such a low wind density, the
shock radius is $\sim 10^{15}$ cm already at 2 days, which means that $\tau_{\rm ff} \sim 3\EE{-7} 
(\Mdot/10^{-9} \msunyr)^2$ at 5.5 GHz at such an early epoch. Free-free absorption is thus insignificant
and can be dismissed from our analysis. We also note that \citet{hor12} used a similar argument to
dismiss free-free absorption in their analysis of radio emission from SN~2011fe. In what follows, we therefore
only consider SSA.

\subsubsection{The wind case, $s=2$.}
\label{sec:wind}

We now compare the radio data for SN~2014J in \S\ref{sec:obs} with the predictions of the model presented in \S\ref{sec:model}.
If the supernova happens in an SD scenario, the accreting WD is expected to have lost some of the accreted material from the donor star 
through a wind. This sets up a $\rho \propto r^{-2}$ circumstellar structure (cf. \S\ref{sec:model}).

As we show below, for the epochs of the radio observations, SN~2014J was clearly in its optically thin phase. This simplifies the expressions above, so that the luminosity becomes
\begin{equation}
	L_{\nu, {\rm thin}}=\frac{8 \pi^{2} kT_{\rm bright} \vartheta_{\nu} r_s^2} {c^2 f\left(\frac{{\nu}_{\rm peak}}{{\nu}_{\rm abs}}\right)}
	                 \nu_{{\rm abs},0}^{(p+3)/2} \nu^{-(p-1)/2},
\label{eq:Luminosity}
\end{equation}
where 

\begin{equation}
	\nu_{{\rm abs},0}= \left(2 \Delta r ~ \varkappa(p) ~N_0 ~ B^{(p+2)/2}\right)^{2/(p+4)}.
\label{eq:nuabs}
\end{equation}
From this, together with expressions in \S\ref{sec:model} and assuming $p=3$, one gets 
\begin{equation}
	L_{\nu, {\rm thin}} \propto T_{\rm bright} ~ \epsilon_{\rm rel}^{1.71} ~ \epsilon_{\rm B}^{1.07}  \left(\dot M / v_w\right)^{1.37} ~ t^{-1.55},
\label{eq:Lum1}
\end{equation}
if $\gamma_{\rm min}$ is not fixed. If it is fixed
\begin{equation}
	L_{\nu, {\rm thin}} \propto T_{\rm bright} ~ \epsilon_{\rm rel}^{0.86} ~ \epsilon_{\rm B}^{1.07}  \left(\dot M / v_w\right)^{1.27} ~ t^{-1.35}.
\label{eq:Lum2}
\end{equation}

At early epochs, when the shock velocity is high,  $\gamma_{\rm min}$ is always larger than unity, and decreases as time goes on. Therefore, Equation \ref{eq:Lum1} applies.
At later epochs, when the shock velocity is such that it would formally imply   $\gamma_{\rm min}  \leq 1$, our constraint on $\gamma_{\rm min}$ takes effect, and Equation \ref{eq:Lum2} applies. 
As stated in \S\ref{sec:model}, we fixed $\epsilon_{\rm rel}$ at 0.1, and allowed  $\epsilon_{\rm B}$ to vary. 
Equations \ref{eq:Lum1} and \ref{eq:Lum2} can be used to scale  $\epsilon_{\rm rel}$ even if only 
$\epsilon_{\rm B}$ is varied. 

In Figure~\ref{fig:Lcurves} we show models for $\epsilon_{\rm B}=0.01$ and 0.1. An almost perfect
overlap between modeled light curves occurs for the combination $\dot M = 7.0\times10^{-10}\msunyr$ 
and $\epsilon_{\rm B} = 0.1$, and $\dot M = 4.2\times10^{-9}\msunyr $and $\epsilon_{\rm B} = 0.01$. Only
at very early epochs ($\lsim 3$ days after explosion), does SSA play a role for the lowest frequencies. The
overlap is not surprising,  since Equation \ref{eq:Lum1} shows that  $\dot{M}/v_{w} \propto \epsilon_{\rm B}^{-0.78}$ for fixed luminosity at early epochs in the optically thin part in our model. We note that \citet{cho12}
 obtain a slightly different power-law index, $-0.7$, in their model. For all models in Figure~\ref{fig:Lcurves},
$\gamma_{\rm min} > 1$ for the time span shown. This means that  Equation \ref{eq:Lum1} describes 
all light curves well, except for the lowest frequencies at $t \lsim 3$ days.

The values of $\dot{M}/v_{w}$ for \sn\ in Figure~\ref{fig:Lcurves} are chosen so that the 5.50 GHz light curves
go through the JVLA 3$\sigma$ upper limit on day 8.2. The light curves for other frequencies lie below
their corresponding upper limits. The second most constraining limit is from our 1.55 GHz eMERLIN 
observation on day 13.8,  yielding $\dot M \lsim 1.15~(7.0) \times10^{-9}\msunyr$ for $\epsilon_{\rm B}= 0.1~(0.01)$ and $v_w = 100~\kms$. We show in Table~\ref{tab:RadioLog}
upper limits for all data points.

Figure~\ref{fig:Lcurves} also includes the most constraining upper limit for SN~2011fe 
\citep{cho12}, together with a 5.9 GHz light curve using $\dot M = 5.0 \times10^{-10}\msunyr$, 
$v_w = 100~\kms$ and $\epsilon_{\rm B}= 0.1$. The limit on mass-loss rate is somewhat below that
of  \citet{cho12}, who obtained $6.0 \times10^{-10}\msunyr (v_w / 100~\kms)$. The difference in those values probably stems from the difference in shell thickness of the emitting region, where we have adopted $\Delta r / r_s = 0.2$  vs. $\Delta r / r_s = 0.1$ \citep{cho12}, and our fixed $T_{\rm bright}$. In any case, the difference
in the upper limit on $\dot{M}/v_{w}$ is much smaller than that  due to the uncertainty in  $\epsilon_{\rm B}$. 

In principle, radio non-detections could also be due to SSA during the observed epochs. In this case,
the observed frequency $\nu_{\rm obs} < \nu_{\rm abs}$, and from Equation \ref{eq:Intensity} we find that the
observed flux at $\nu_{\rm obs}$ is $\propto r_s^2 (v_{\rm obs}^5/v_{\rm abs})^{1/2}$. The combination of
$ r_s^2 v_{\rm abs}^{-1/2}$ is a weak function of $\dot{M}/v_{w}$, and to make SSA important for the
observations discussed here would require  values of $\dot{M}/v_{w}$ much larger than those at which free-free
absorption becomes important. We can therefore fully dismiss SSA as a cause for the radio non-detections of SN~2014J.

\begin{figure}
\centering
\includegraphics[width=9cm,angle=0]{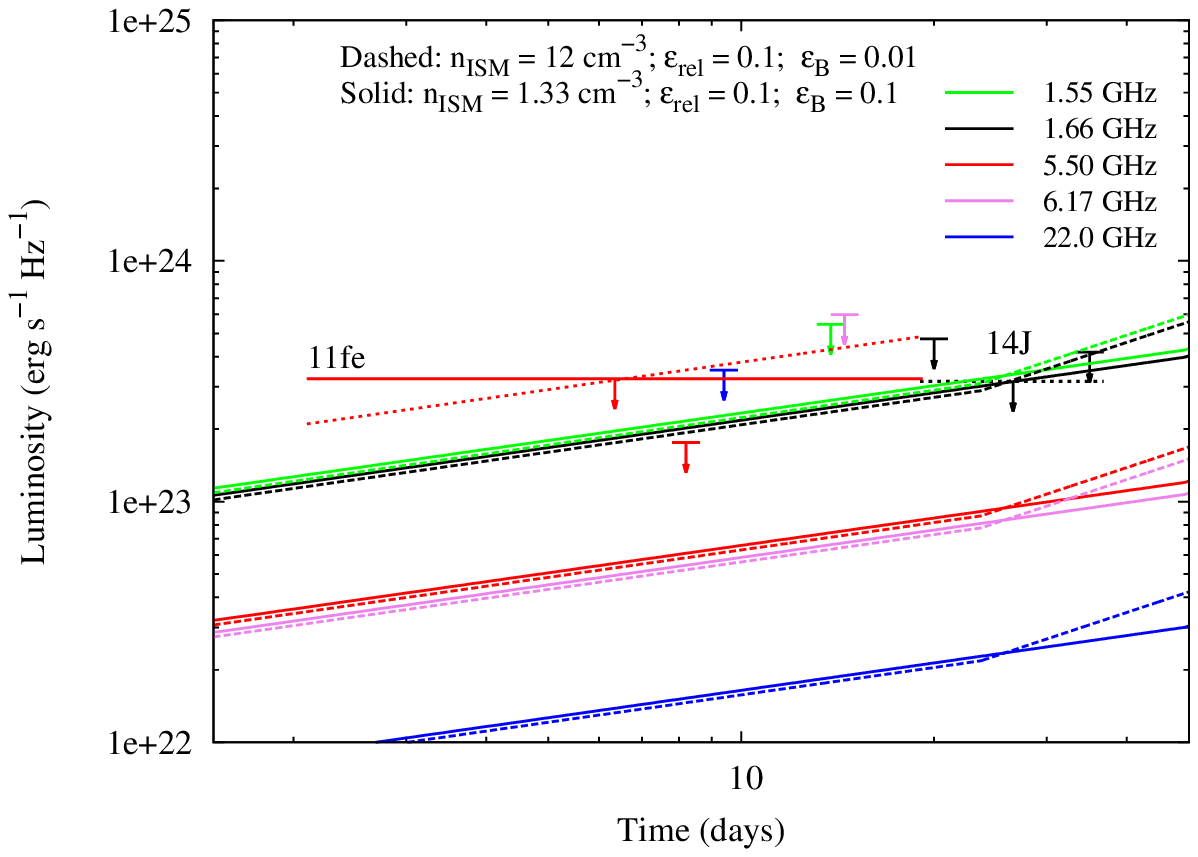}
\caption{
Same as in Figure~\ref{fig:Lcurves}, but for circumstellar gas with constant density ($s = 0$),  i.e., the insterstellar medium (ISM). Solid lines are for  $n_{\rm ISM} = 1.33 \cm3$, and dashed line is for $n_{\rm ISM} = 12 \cm3$, assuming 
$\epsilon_{\rm B} = 0.1$ and $\epsilon_{\rm B} = 0.01$, respectively. The data upper limits for \sn\ are the
same as in Figure~\ref{fig:Lcurves}. The constraint on $n_{\rm ISM}$ is set by the stacked 1.66 GHz eEVN 
data from days 20 and 35. The change in spectral slope around 25 days for the $\epsilon_{\rm B} = 0.01$ 
model is due to the $\gamma_{\rm min}  \geq 1$ constraint then coming into effect. For a comparison we 
also show the 3$\sigma$ upper limit from the stacked 5.9 GHz sample for SN~2011fe between days 
$2.1-19.2$ \citep{cho12}, together with part of the light curve for a model (dotted red line) 
assuming $n_{\rm ISM} = 7.0 \cm3$ and $\epsilon_{\rm B} = 0.1$.
}
\label{fig:Lcurves_constant}
\end{figure}

\begin{figure}
\centering
\includegraphics[width=9cm,angle=0]{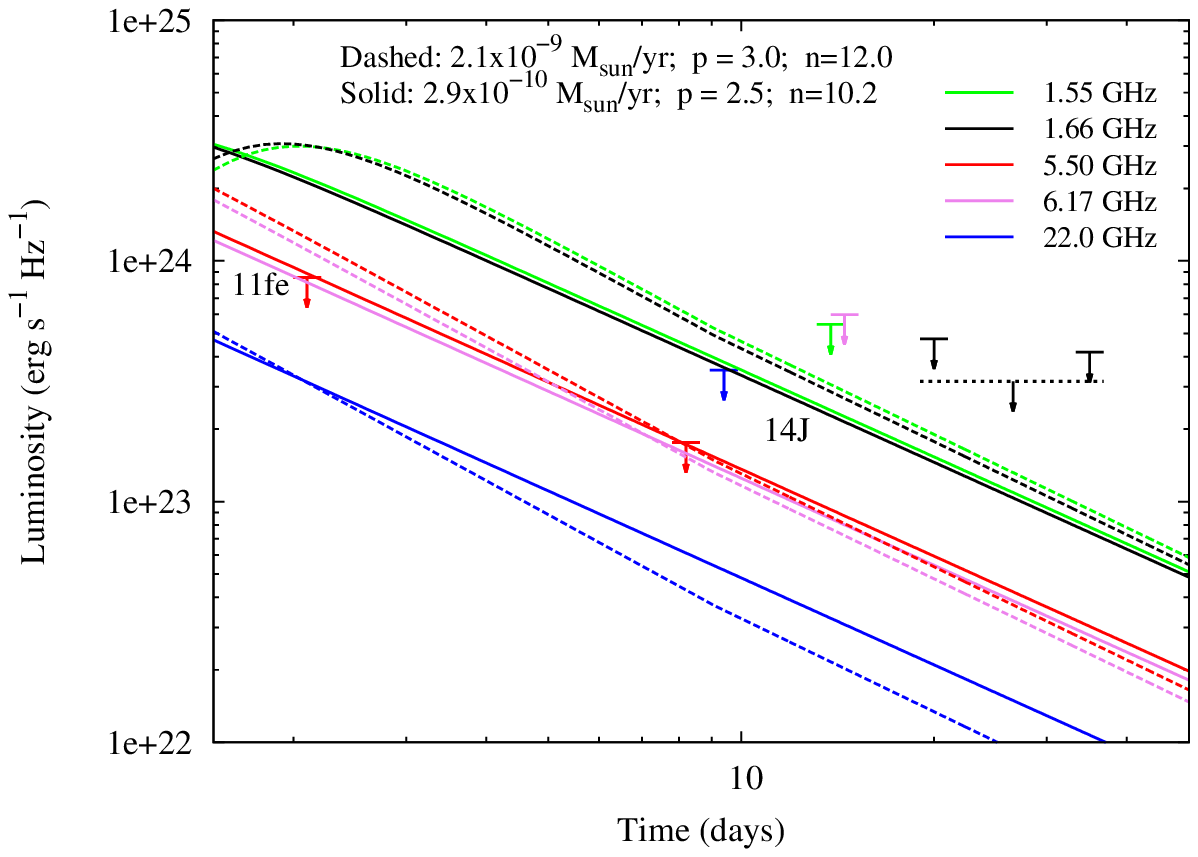}
 \caption{
Same as in Figure~\ref{fig:Lcurves}, but for different values of $p$ and $n$. The values of $\dot{M}/v_{w}$
were chosen for the 5.50 GHz light curve to cross the upper JVLA limit from day 8.2. Note the change in
slopes for the dashed lines around day 9, when the condition $\gamma_{\rm min} \geq 1$ comes into
effect. See text for further details.}
\label{fig:Lcurves_params}
\end{figure}

\subsubsection{The constant density case, $s=0$.}
\label{sec:ism}

If the progenitor  of \sn\ followed the double-degenerate channel, 
then the exploding WD is expected to be surrounded by the interstellar medium (ISM),  which has a constant density. \citet{cho12} discussed this scenario for SN~2011fe, and obtained a limit for the density of
$\lsim 6 \cm3$ ($\epsilon_{\rm B} = 0.1$).

The general behavior of the radio light curves for the constant density ISM case is different from the 
$n_{\rm CSM} \propto r^{-2}$ wind case in \S\ref{sec:wind}. For the constant density case and $p=3$, the radio 
luminosity increases with time \citep[see also][]{cho12} according to
\begin{equation}
	L_{\nu, {\rm thin}} \propto T_{\rm bright} ~ \epsilon_{\rm rel}^{1.71} ~ \epsilon_{\rm B}^{1.07}  ~ n_{\rm ISM}^{1.10} ~ t^{0.38},
\label{eq:Lum3}
\end{equation}
if $\gamma_{\rm min}$ is not fixed. If it is fixed
\begin{equation}
	L_{\nu, {\rm thin}} \propto T_{\rm bright} ~ \epsilon_{\rm rel}^{0.86} ~ \epsilon_{\rm B}^{1.07} ~ n_{\rm ISM}^{1.27} ~ t^{0.88}.
\label{eq:Lum4}
\end{equation}

Here, we substituted $n_{\rm CSM}$ with $n_{\rm ISM}$ to highlight the likely origin of the gas in 
the $s=0$ case. Figure~\ref{fig:Lcurves_constant} shows models with densities 
$n_{\rm ISM} = 1.33 \cm3$ 
($\epsilon_{\rm B} = 0.1$) and $n_{\rm ISM} = 12 \cm3$ ($\epsilon_{\rm B} = 0.01$). Scaling according 
to $n_{\rm ISM} \propto \epsilon_{\rm B}^{-0.97}$ (cf. Equation \ref{eq:Lum3}) makes the light curves for 
these models overlap fully, except for $t \gsim 25$ days, when the condition 
$\gamma_{\rm min}  \geq 1$ becomes important for the $n_{\rm ISM} = 12 \cm3$ model. The most
constraining data are our eEVN 1.66 GHz data, stacked together, and the model parameters were 
chosen for the modeled radio luminosity to match those data. However, due to the 
$\gamma_{\rm min}  \geq 1$ constraint, the 35 day data alone are almost as constraining as the
stacked data.

In Figure~\ref{fig:Lcurves_constant}, we 
also show the stacked 5.9 GHz data for SN~2011fe \citep{cho12}, together with a model characterized by 
$n_{\rm ISM} = 7.0 \cm3$ and $\epsilon_{\rm B} = 0.1$. We are thus close to \citet{cho12} 
regarding the limit on $n_{\rm ISM}$ for SN~2011fe. For $s=0$, we  used $\Delta r / r_s = 0.12$,
which is close to the value 0.1 used by \citet{cho12}. The limit on $n_{\rm ISM}$ we find 
for \sn\ is $\approx 5.3$ times lower than for SN~2011fe, and is therefore clearly the lowest limit on density 
for the constant density case in any SN~Ia. 

As for the $s=2$ case, SSA is unimportant for the $s=0$ case. To be efficient enough to mute the
radio emission to be consistent with the observed upper limit,  $n_{\rm ISM}$ would have to be
$\gsim 10^{14} \cm3$ (for $\epsilon_{\rm B} \leq 0.1$), which is fully ruled out from X-ray limits, as well the 
normal optical behaviour of the supernova.

\subsection{Sensitivity of results to parameters}
\label{sec:sens-res}

While usually not acknowledged in the SN Ia literature,  the obtained radio and X-ray upper limits on the circumstellar density  are model dependent. In particular, Equations \ref{eq:Lum1}  through \ref{eq:Lum4} show how most parameters influence the 
results. 

We already mentioned the uncertainty in $\epsilon_{\rm rel}$ and, especially, $\epsilon_{\rm B}$,
whereas $T_{\rm bright}$ is observationally constrained by other similar radio sources.  The thickness
of the radio-emitting region is yet another source of uncertainty, but probably small in comparison to other
uncertainties.

There is also an uncertainty in the upper limit on $\dot{M}/v_{w}$ due to values chosen for $n$ and $p$. 
Figure~\ref{fig:Lcurves_params} shows two models, both with $\epsilon_{\rm B} = 0.1$ and 
$\epsilon_{\rm rel} = 0.1$, but where we have also assumed $p=2.5$ and $n=10.2$ 
(solid lines), and $p=3$ and $n=12$ (dashed lines). In both models  we have used the earliest $3\sigma$ 
JVLA point for \sn\ to constrain $\dot{M}/v_{w}$. For the $p=2.5$ model, 
$\dot{M}/v_{w} = 2.9\EE{-10} \msunyr$, i.e., a factor $\approx 2.4$ lower than the $p=3$ model in 
Figure~\ref{fig:Lcurves}. The choice of $p=3$ therefore gives
a conservative limit on $\dot{M}/v_{w}$, unless $p>3$. Judging from Type Ib/Ic SNe,
the expected deviation from $p=3$ is not large \citep{che06}, so we do not consider the uncertainty in 
$p$ being a major source of uncertainty for $\dot{M}/v_{w}$.

For the $n=12$ model, 
$\dot{M}/v_{w} = 2.1\EE{-9} \msunyr$, i.e., a factor $\approx 3.0$ higher than the $n=10.2$ model in 
Figure~\ref{fig:Lcurves}. The choice of $n=10.2$ may give too low a limit on $\dot{M}/v_{w}$, unless 
a  shallower density profile than $n=10.2$ is expected. Indeed, \citet{dwa98} argue that an exponential
density profile of the outer ejecta fits early explosion models better than a power law, and could
indicate steeper profiles than at least $n=7$ for the outermost ejecta. We have also run models
for $n=7$, but the value for $\dot{M}/v_{w}$ then becomes so low ($< 10^{-11} \msunyr$) 
that the model breaks down, producing shock velocities in excess of $c$. For such low wind densities,
a relativistic treatment of the shock interaction is needed, similar to for gamma-ray bursts. From the models  of \citet{matzner99} it seems reasonable to assume that $n \gsim 9$ for the outermost ejecta, which in our  model would imply an upper limit of $\dot{M}/v_{w} \approx 2.0 \EE{-10} \msunyr$. The span in upper limit on $\dot{M}/v_{w}$ by a factor of $\approx 10.5$ between $9\leq n \leq 12$ shows that the unknown density profile for the outermost ejecta is an important source of uncertainty, and that accurate models for the outermost ejecta are needed. This is even more evident for the constant density case $s=0$, for which we find that 
our limit on $n_{\rm ISM}$ ranges from $\approx 0.17 \cm3$ for $n=9$ to $\approx 6.3 \cm3$ 
for $n=12$, assuming $\epsilon_{\rm B} = 0.1$, $\epsilon_{\rm rel} = 0.1$ and $p=3$. The solution 
for $n=9$ is, however, unphysical due to too large velocities for the shock during the first $\sim 10$ days, 
calling for a relativistic treatment of the dynamics.

\section{Discussion}

\subsection{The possible progenitors of SN~2014J}

\subsubsection{Single Degenerate progenitor systems}
\label{sec:sd}

The SD progenitor systems involve only one WD  and include, in decreasing order of  mass-loss rate from the supernova progenitor, symbiotic systems, WDs with steady nuclear burning, and recurrent novae.

\begin{figure*}
\centering
\includegraphics[width=19.5cm,angle=0]{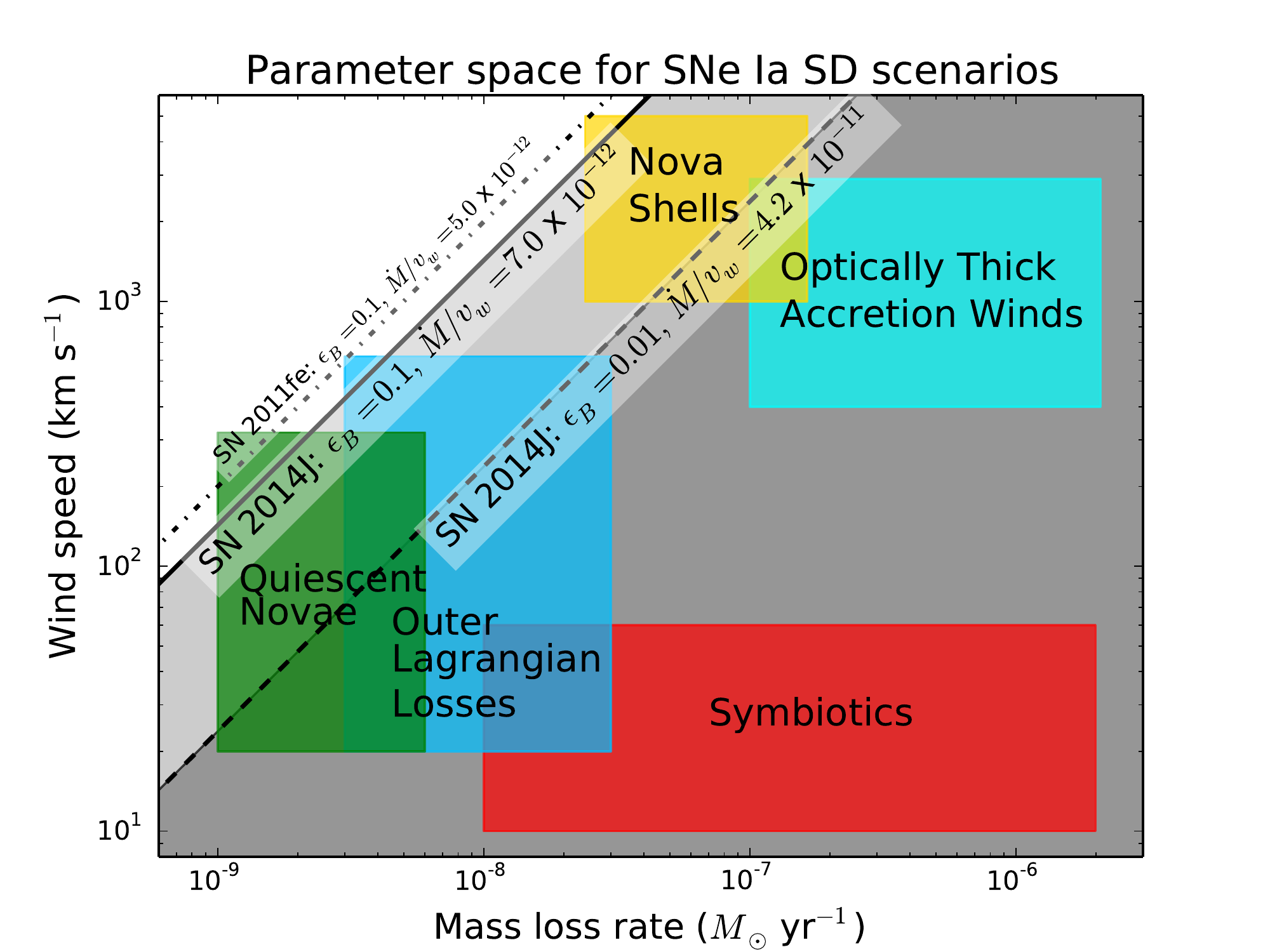}
\caption{
Constraints on the parameter space (wind speed vs. mass-loss rate) for single degenerate scenarios for SN~2014J. The progenitor scenarios discussed in \S\ref{sec:sd} are plotted as schematic zones, following \citet{cho12}.
We indicate our 3$\sigma$ limits on $\mdot/v_{w}$, assuming $\epsilon_{\rm B}$ = 0.1 
(solid; corresponding to the solid curves in Figure~\ref{fig:Lcurves}) and the conservative case of 
$\epsilon_{\rm B}$ = 0.01 (dashed; corresponding to the dashed curves in Figure~\ref{fig:Lcurves}).  
Mass loss scenarios falling into the gray shaded areas should have been detected 
by our deep radio observations, and therefore are ruled out for SN~2014J. Not included in the figure is
the spun-up/spun-down progenitor scenario discussed in \S\ref{sec:broad}, as this predicts a constant density. For
a comparison, we include our reassessed limit for SN~2011fe (dash-dotted line) for the same choice of parameters as the solid line for SN~2014J, which essentially leaves only room for quiescent nova emission as a viable alternative among the SD scenarios for SN~2011fe.
}
\label{fig:mdot-vwind}
\end{figure*}

In a symbiotic system, the WD accretes mass from a giant star \citep{hachisu99}. The WD loses this 
accreted matter at rates of $\dot{M} \gtrsim 10^{-8}$\msunyr and $v_w\approx 30$\kms. The radio emission from those systems should have been detected by our deep sensitive observations. Thus, our
radio non-detection rules out a symbiotic system as the progenitor of SN~2014J 
(red region in Figure \ref{fig:mdot-vwind}).

Another possible SD scenario is one where  a main sequence, subgiant, helium, or giant star
undergoes Roche lobe overflow onto the WD at rates of  $3.1\times 10^{-7} \msunyr \lesssim \dot{M} \lesssim 6.7\times 10^{-7} \msunyr $ 
\citep{nomoto07}. At those accretion rates, the WD experiences steady nuclear burning \citep{shen07}. 
For an assumed fraction  $\epsilon_{\rm loss} = 0.01$ of the transferred mass to be lost from the system,  
the mass-loss rate is constrained to $3.1\times 10^{-9} ~\msunyr  \lesssim \dot{M} \lesssim 6.7\times 10^{-9} ~\msunyr$  and  typical speeds of  100 \kms $\lesssim v_w \lesssim$ 3000 \kms, 
where the low speeds apply for steady nuclear burning, 
while the high speeds apply to the systems with the highest accretion rates. 
At the lower end of $\mdot$, the mass loss through the outer Lagrangian points of the system proceeds at speeds up to $\sim$600 \kms.
 Most of the parameter space for the low-accretion rate scenario is ruled out by our radio observations, 
 if $\epsilon_{\rm {\rm B}} \simeq 0.1$ (blue region in Fig. \ref{fig:mdot-vwind}).
 At the upper end of $\mdot$ the winds become optically thick, limiting the accretion rate to $\dot{M}_{\rm acc} \approx 6\times 10^{-7} ~M_{\odot}~\rm yr^{-1}$ and wind speeds of a few $\times $ 1000~\kms \citep{hachisu99,hachisu08}.  Our data  essentially rule out completely the high-accretion rate scenario of a WD with steady nuclear burning (cyan region in Fig. \ref{fig:mdot-vwind}).

Finally, another  possible SD channel is that of recurrent novae, which lie at the lowest accretion rate regime among popular SD scenarios. Here, a WD accreting at a rate  
$\dot{M} \approx (1-3) \times10^{-7}~\msunyr$,
ejects  shells of material at speeds of a few $\times1000 \kms$, with
typical recurrence times of a few years.
The radio observations in Table \ref{tab:RadioLog} probe a radius of $\simeq (0.7-2.6) \times 10^{16}$~cm
(for $s=2$ and $\epsilon_{\rm B}=0.1$), 
which constrains the presence of shells with recurrence times of 
$\lesssim 1.6~(v_{\rm shell}/2000~\kms)^{-1}\ (r_{\rm shell}/10^{16}\ {\rm cm})$~yr.
Models of recurrent novae  seem to indicate that as much as $\sim$15\% of the accreted material over the recurrence time is ejected \citep{yaron2005,shen2009}. For the typical accretion rates above, this implies an ejected shell mass of 
$\approx (2.4-7.1) \times10^{-8} \ (v_{\rm shell}/2000~\kms)^{-1}\ (r_{\rm shell}/10^{16}\ {\rm cm})~M_{\odot}$, 
which should have been detected by our sensitive observations (see gold region in Fig. \ref{fig:mdot-vwind}). Unfortunately, the short duration of the nova radio burst, a few days at most, may have prevented its detection, so we cannot  rule out completely the possibility of a nova shell ejection.
During the quiescent phase between nova shell ejections, the WD accretes at a rate of 
$\dot{M} \sim 1 \times10^{-7}~\msunyr$, so that the mass-loss wind parameter  
is $\dot{M}/v_w \sim 1 \times10^{-9}\ (\epsilon_{\rm loss}/0.01)/100$\kms. 
If $\epsilon_{\rm B} = 0.1$, our observations rule out almost completely the scenario 
with WD accretion during the quiescent phase of the star,  
 whereas the case with $\epsilon_{\rm B} = 0.01$ cannot be excluded completely (green region in Fig. \ref{fig:mdot-vwind}).

In summary, our observations exclude completely symbiotic systems and the majority of the parameter space associated with stable nuclear burning WDs, as viable progenitor systems for SN~2014J. Recurrent novae with main sequence or subgiant donors cannot be ruled out completely, yet most of their parameter space is also excluded by our observations. 

\subsubsection{Double Degenerate progenitor systems}
\label{sec:dd}

The alternative to a SD scenario is the DD channel, which involves two WDs in a binary system.
In this case, the progenitor  star is expected to have exploded in a constant ambient density medium. 
We can estimate the density in the region surrounding SN~2014J from the 
column density of neutral hydrogen toward the supernova position, $N_{\rm H\,I} \sim 2 \times 10^{20}$ cm$^{-2}$ \citep{zwa08}. Assuming  a path length, $l \sim$ 100 pc, and  solar abundance ($\mu \approx 1.4$), the particle number density at the supernova location is $\mu\,N_{\rm H\,I} / l \sim$ 0.9 cm$^{-3}$. Our stacked eEVN limits imply $n_{\rm ISM} \lesssim 1.3 (12)$ cm$^{-3}$ for $\epsilon_B = 0.1$ (0.01), and are thus consistent with the SN directly expanding into the interstellar medium. Therefore, our radio non-detections  
are consistent with the DD channel for SNe~Ia.

We note that the limit  imposed on $n_{\rm ISM}$ by the microphysical parameter $\epsilon_{\rm B} = 0.1$ is formally similar to the likely value of $n_{\rm ISM}$ at the SN location. Yet, the uncertainties involved in this estimate are such that both values are in agreement. At any rate, the sensitivity of $\epsilon_{\rm B}$ on
$n_{\rm ISM}$ demonstrates the usefulness of late-time radio observations to constrain this relevant 
microphysical parameter in SNe~Ia. For example, a non-detection of \sn\ one year after explosion
with the same observational limit as from our stacked 1.66 GHz observations would, according to our model,
constrain $n_{\rm ISM}$ to $\lsim 0.27 \cm3$, assuming $\epsilon_{\rm {\rm B}}=0.1$, $\epsilon_{\rm rel}=0.1$, $n=10.2$, $p=3$ and $\gamma_{\rm min}=1$. The supernova shock wave will at this point be located 
at $r_s \approx 1.6\EE{17}$ cm. A non-detection at such late epochs and such low flux levels, will 
certainly be very useful in constraining $\epsilon_{\rm B}$.

\subsection{Broader picture}
\label{sec:broad}

The recent and nearby SNe 2011fe and 2014J have offered a remarkable possibility to learn
about the origin of SNe Ia. We have shown here that deep radio observations can be used to rule
out several progenitor models for \sn\, and a similar discussion was made for SN~2011fe by \citet{cho12}.
In addition to this, \citet{mar12} provided deep limits for SN~2011fe from X-rays (cf. \S\ref{sec:intro}), which do not 
depend on $\epsilon_{\rm B}$, but where the limit on circumstellar density has a stronger dependence on 
$\epsilon_{\rm rel}$ and $\gamma_{\rm min}$ than for radio emission. While $\epsilon_{\rm rel} =0.1$ has
 been used by us and others, it must be cautioned that, e.g., SN~1993J had a much lower value \citep{fra98,per01,vidal11}, although it is clear that this  supernova bears little resemblance with Type Ib/Ic SNe, which have been used as templates to model SNe Ia.

For SN~2011fe, the non-detections in radio and X-rays were accompanied with no circumstellar line 
absorption \citep{pat13} and a non-detection of late nebular emission from gas ablated off an SD companion \citep{sha13b}. This, together with other evidence for SN~2011fe \citep[see][]{mao14}, has been used to 
argue for an increased likelihood of SNe Ia being the endpoint of a DD scenario rather than SD scenarios. 
Our non-detections of radio emission from \sn\ in principle add to this evidence. 

However, \citet{jus11} suggested that a SD scenario, with a spun-up/spun-down super-Chandrasekhar WD  \citep[see also][]{dis11,hachisu12}, can still be possible if the donor star shrinks far inside its Roche lobe prior to the explosion. This would make the SD companion smaller and more tightly bound, and only very dilute circumstellar gas would be expected in the immediate vicinity of the WD.  \citet{dis11} argue that density could be of the same order as typical interstellar 
densities. Continued radio observations of both SNe~2011fe and 2014J could be useful to test the presence
of such low-density gas (cf. \S\ref{sec:dd}).

For a typical time-scale of $\sim 10^3$ years between last Roche-lobe overflow and explosion \citep{jus11}, and a wind speed of $100 \kms$, the last traces of substantial circumstellar gas could in this scenario be at 
a distance of  $\sim 3\EE{17}$ cm, and could explain the presumed shell around, e.g., SN~2006X 
\citep{pat07}. If the supernova ejecta would start to interact with such a shell, radio emission would increase. 
Radio observations of SN~2006X \citep{cha08} two years after explosion, however, failed to detect any 
emission. \citet{pat07} estimated a shell radius of $\sim 10^{16}$~cm, which according to the estimate 
in \S\ref{sec:sd}, was most likely overtaken by the supernova ejecta by $t \approx 2$ years. Continued monitoring of \sn\ would be useful to trace these putative shells.

We emphasize that shells around supernovae do not have to lie
along the line of sight to be detected in radio, as opposed to the narrow absorption line features. This
should increase the possibility to detect such shells in the radio, especially if shells are as common as
suggested (see \S\ref{sec:obs}).  Interaction with shells are also better observed in the radio than 
in X-rays, as no inverse Compton scattering is expected at late times when the supernova has faded
and the distance from the line emitting SN ejecta and the shell is large.

\citet{sha13a} caution that the lack of signatures from an SD companion could be a problem for the model of \citet{jus11}, as only $\lsim 0.001~\msun$ of ablated mass from the companion can be accommodated
by the observations of \citet{sha13b}, in combination with an extrapolation of the models presented in \citet{mat05} and \citet{lun13}, before giving rise to detectable H$\alpha$ emission in the nebular phase; all models calculated by \citet{mar00}, \citet{pan12} and \citet{liu12} predict more than $\sim 0.01~\msun$ of 
ablated mass.  
A way to avoid H$\alpha$ emission is, of course, if the donor is He-rich. The models of \citet{liu13} show that
of order $\sim 0.02~\msun$ of He-rich gas would then reside in the centre of the supernova in the nebular
phase. However, even if the donor is H-rich, the caution by \citet{sha13a} should not be over interpreted.
 
The opacity in the nebular emission models of \citet{mat05} and \citet{lun13} does not contain as many
spectral lines as more recent models by, e.g., \citet{jer11}. Scattering in the spectral region around 
H$\alpha$ could be more severe than previously anticipated, and the constraint from lack of nebular 
H$\alpha$ less important. Nebular lines further out in the red should suffer from less scattering, and the 
models discussed in \citet{lun13} show that narrow ($\lsim 10^3 \kms$) [Ca~II] lines are present in the red, 
and could thus be more constraining than H$\alpha$. Traces of these lines could be useful to test scenarios with both H-  and He-rich donors. Further detailed modeling of the nebular phase is indeed needed, as well as modeling of supernova ejecta colliding with compact companions such as  those in the models of \citet{jus11}.

Deep nebular spectra of SN~2014J are warranted to test those impact models, due to the SN proximity.
A potential problem  is, however, the extinction toward  SN~2014J and the  
complicated interstellar imprint on the supernova spectrum \citep{goo14,Welty2014}. 
The latter may, in particular, make the search for narrow-line variations more cumbersome than for, e.g., SN 2006X. 

\subsection{Future outlook for radio observations of SNe Ia}
At the moment, our deepest radio limits on circumstellar gas are for SNe~2011fe and 2014J. With the advent of the
Square Kilometre Array (SKA), we will be able to obtain significantly deeper radio  limits (or, potentially, a detection) for 
SNe Ia exploding at the distance of M~82.  For more distant supernovae, we will obtain similar limits to those obtained
for SNe~2011fe and 2014J, which will allow us to build a picture from a larger statistical sample. 

The first phase of SKA considers three different components. One of them, SKA1-mid, promises to yield 1$\sigma$ sensitivities of $\sim 0.7 \mu$Jy/b in one hour at a fiducial frequency of 1.7 GHz. This figure is five times better than currently provided by the most sensitive array, the JVLA. Therefore, SKA1-mid should be able to either detect the putative radio emission of SN~2014J-like objects up to distances $\lesssim 8$ Mpc in less than one hour, or put significantly better constraints on some of the parameter space of SD scenarios for the next SN Ia that explodes in M~82, some of which could not be completely ruled out even by our very deep radio observations.
However, the expected number of SN~Ia per year in such a volume of the local universe is small. Indeed, 
since the volumetric SN Ia rate is $\sim 3\times 10^{-5}$ SN/yr/Mpc$^{-3}$ \citep{dilday10}, 
we should expect on average one SN Ia every $\sim$15 yr within a distance of $\lsim$8 Mpc, which is a 
small value. To obtain a statistically significant sample of SNe~Ia observed in radio, with similar
upper limits to those obtained by us for SN~2014J, we need to sample significantly larger volumes and need
much more sensitive radio observations. For example,  by sampling out to a distance of 25 Mpc, we can 
expect $\sim$2 SNe~Ia per year within the sampled volume, which in 10 years would result in a total of $\sim$20 SNe~Ia,
enough to extract statistical results.  At this maximum distance, we need a sensitivity of $\sim$50 times better
than obtained by the observations discussed here, to be as constraining, or $\sim 80$ nJy/b. 
When SKA is completed, the fiducial 1$\sigma$ sensitivity should be 10 times better than for SKA1-mid, 
or about $\sim 70$~nJy/b in one hour, and such statistical studies will be perfectly possible in short amounts 
of time. At this level of sensitivity, a non-detection would be essentially as meaningful as a direct 
detection, since the former would imply that only the DD scenario is viable, while the latter would tell us which 
of the SD channels result in SNe Ia.

\section{Summary}

We report deep eEVN and eMERLIN radio observations of the Type Ia SN~2014J
in the nearby galaxy M~82, along with a detailed modeling of its radio emission.
Our observations result in non-detections of the radio emission from SN~2014J. Yet, 
radio data and modeling allow us to place a tight constraint on the mass 
loss rate from the progenitor system of SN~2014J. Namely, if the exploding WD was surrounded by a
wind with a density profile $\rho \propto r^{-2}$, as expected for a SD scenario, then our upper limit to the mass-loss rate is  
$\dot{M} \lesssim 7.0\times 10^{-10}~\msunyr$, for a wind speed of $100 \kms$.

If, on the contrary,  the circumstellar gas has a constant density, as 
expected to be the case for the DD scenario (but also in a small region of the parameter space of  SD scenarios), then our modeling yields an upper limit on the gas density, such that $n_{\rm ISM} \lesssim 1.3 \cm3$.

Our stringent upper limits to the circumstellar density around SN~2014J allow us to exclude completely symbiotic systems and the majority of the parameter space associated with stable nuclear burning WDs, as viable progenitor systems for SN~2014J. 
For the case of recurrent novae with main sequence or subgiant donors, we cannot rule out them completely, yet most of their parameter space is also excluded by our observations
for the standard assumption of $\epsilon_{\rm B} = 0.1$, where $\epsilon_{\rm B}$ is the
ratio of magnetic energy density to post-shock thermal energy density.

We have also reassessed the radio limits on wind density for SN~2011fe, and for $\epsilon_{\rm B} = 0.1$ we 
obtain $\dot{M} \lesssim 5.0\times 10^{-10}~\msunyr$ (for a wind speed of $100 \kms$) and 
$n_{\rm ISM} \lesssim 7.0 \cm3$. These limits are close to those calculated by \citet{cho12}. Our limit on 
$\dot{M}/v_w$ for \sn\ is thus similar to that for SN~2011fe, whereas for the constant density case 
we obtain a much lower limit than for SN~2011fe, and hence the lowest limit for a constant density 
ambient around a SN~Ia.

The combined radio limits on circumstellar gas around SNe~2011fe and 2014J add to evidence from mainly non-detections of X-rays from SN~2011fe  and 2014J\citep{mar12,mar14} and no detection of H$\alpha$ in the nebular
phase of SN~2011fe \citep{sha13b}, that SNe~Ia are very likely to stem from the DD scenario, rather than SD scenarios. 

Finally, we highlight future observations with the Square Kilometre Array (SKA). When fully completed, the SKA is 
likely to yield limits on circumstellar gas for future SNe~Ia similar to the limits reported here for the nearby SNe~2011fe and 2014J, but for distances well beyond the Virgo cluster. For nearby SNe~Ia, SKA limits are 
likely to be fully conclusive regarding the origin of the progenitor systems of SNe Ia.

\acknowledgments

We are grateful to Carles Badenes for useful comments on the manuscript, to Rub\'en Herrero-Illana for Pythonic advice to produce Figure 5, and to Vicent Peris and Oscar Brevi\`a from the Observatorio Astron\'omico de Aras (Valencia, Spain) for the optical image of M82 used in Figure 1.
We acknowledge the eMERLIN and EVN programme committees, and the directors of the EVN stations, for supporting the radio  observations of SN\,2014J.
The European VLBI Network (EVN) is a joint facility of European, Chinese, South African, and 
other radio astronomy institutes funded by their national research councils. 
The electronic Multi-Element Radio Linked Interferometer Network (eMERLIN) is 
the UK's facility for high resolution radio astronomy observations, operated by The University of Manchester for the Science and Technology Facilities Council (STFC).
The research leading to these results has received funding from the European Commission Seventh Framework Programme (FP/2007-2013) under grant agreement No 283393 (RadioNet3).
AA, JCG, JMM, MAPT, ER, and IMV acknowledge support from the Spanish MICINN through grants AYA2012-38491-C02-01 and AYA2012-38491-C02-02. 
P.L. acknowledges support from the Swedish Research Council.
The research leading to these results has received funding from the European Commission Seventh Framework Programme (FP/2007-2013) under grant agreement No 283393 (RadioNet3).

{\it Facilities:} \facility{eEVN}, \facility{eMERLIN}.

\end{document}